\documentclass[twocolumn,aps,prb,floatfix]{revtex4}

\usepackage{graphicx}
\usepackage{epstopdf}
\usepackage{amsmath}
\usepackage{latexsym}
\usepackage{amsmath}

\newcommand{\beq}{\begin{equation}}
\newcommand{\eeq}{\end{equation}}

\begin{document}

\title{Electronic and optical gap renormalization in carbon nanotubes near a metallic surface}

\author{Catalin D. Spataru}
\email{cdspata@sandia.gov}
\affiliation{Sandia National Laboratories, Livermore, California 94551, USA}

\begin{abstract}
Renormalization of quasiparticles and excitons in carbon nanotubes (CNTs) near a
metallic surface has been studied within a many-body formalism using an embedding approach newly implemented in the GW and Bethe-Salpeter methods. The quasiparticle
bandgap renormalization in semiconducting CNTs is found to scale as
$-1/(2h_a)$, with $h_a$ the apparent nanotube height, and it can exceed half an eV. 
Also, the binding energy of excitons is reduced dramatically -by as much as $75\%$- near the surface. Compensation between quasiparticle and excitonic effects results in small changes in the optical gap. 
The important role played by the nanotube screening response in establishing these effects is emphasized and a simple electrostatic model with no adjustable parameters explains the results of state-of-the-art calculations and generalizes them to a large variety of CNTs.
\end{abstract}

\maketitle

\section{Introduction}
Carbon nanotubes (CNTs) have attracted tremendous interest for a number of applications, and also for the breadth of new scientific questions that they bring. The properties of isolated CNTs have been extensively studied. In most situations, however, these systems are subject to external perturbations, which make their behavior deviate from the ideal, isolated case. Examples include interaction with a substrate, other nanostructures, polymers or DNA encapsulation, metallic contacts, doping, applied electric or magnetic fields, applied strain, alignment in periodic arrays, and so on \cite{CNTbook}. 
Both electronic and optical properties of CNTs are expected to be altered by such environmental and dimensionality effects \cite{Spataru_dop,Rohlfing_2tubes,Ando_env,SpataruFLstrain}. 

Predicting the changes in properties of CNTs due to a substrate is  important not only for the potential integration of CNTs in functional devices but also from a fundamental physics perspective. Consider a semiconducting CNT in a field-effect transistor configuration. The alignment of electronic states at the metal/CNT interface is critical to the device performance as it determines the activation energy necessary to inject an electron from the metal contact \cite{Leonard}. How do the metal contacts or even the metallic gate affect the electronic states of the CNT? 

Recent experimental measurements of CNTs on a Au substrate find that the quasiparticle (QP) bandgaps of semiconducting CNTs are renormalized appreciably by the substrate, even for a nanotube-substrate separation as large as $1$ nm \cite{Lin}. 
The binding energy of excitons ${ E_b}$ is another fundamental quantity important in optoelectronic and photonic applications, as it directly affects critical parameters such as exciton dissociation, electron-hole recombination or radiative decay rates. In isolated semiconducting CNTs excitons bind with energies amounting to a large fraction of the QP bandgap \cite{Wang,Lin}. One question is whether ${ E_b}$ is also significantly altered near a metal surface.

In this work I study the energy renormalization of QP and excitons in semiconducting CNTs near a metallic surface. I consider nanotube-surface separations ranging from weak physisorption all the way to the isolated nanotube case. For these separations hybridization and charge rearrangements effects \cite{Zojer} at the interface between the CNT and the metallic surface can be neglected. Large separations can be realized experimentally via a thin insulating spacer \cite{Hong,Sakashita}, which may serve to prevent charge-transfer effects. Charge-transfer can take place in CNTs contacted by a metal surface \cite{Clair}; however, in devices with top-contact geometry,
 the CNT-doping level is controlled by the gate   \cite{Cummings}. Thus, in relevant applications charge-transfer effects can be disentangled from other effects and are dropped from this work having considered them elsewhere \cite{Spataru_dop,Spataru_dop2}.

\section{Many-body formalism}
The theoretical approach is based on \textit{ab initio} methods that take into account many-electron correlation effects known to play an important role in the electronic and optical properties of CNTs \cite{Spataru,Ando,Chang,Perebeinos,Zhao}, specifically the ${ G_0W_0}$ approximation \cite{HL} for the electron self-energy, and the Bethe-Salpeter (${ BS}$) equation \cite{Rohlfing} for excitonic effects. 
The many-body Green's function approach involves a perturbation expansion, to first order, about the screened Coulomb interaction  ${ W=\epsilon^{-1}V}$, where ${ V}$ is the bare Coulomb interaction. The dielectric function ${ \epsilon=1-VP}$ is evaluated from the irreducible polarizability ${ P}$ within the Random Phase Approximation (RPA). The one-particle Green's function ${ G}$  is constructed from Kohn-Sham eigenvalues and eigenfunctions obtained within Density Functional Theory (DFT) in the Local Density Approximation (LDA) \cite{KS} using the Quantum ESPRESSO package. \cite{QEspresso}

Commensurability constraints and large nanotube-surface separations make full \textit{ab initio} many-body calculations computationally prohibitive. To make calculations feasible, it is advisable to take advantage of the absence of hybridization effects and use an embedding approach that enables many-body {\it ab initio} calculations on the nanotubes after integrating out the electronic degrees of freedom of the surface. 
This approach can be applied generally to the study of QPs and excitons of various nanostructures in different complex environments, as long as the electronic ground-state properties of the nanostructure are not significantly altered by the surrounding environment. 

\subsection{Embedding approach for $GW$ and $BS$ calculations}

To begin, note that the absence of hybridization effects leads within LDA to substrate-independent nanotube states. The RPA irreducible polarizability ${ P}$ 
of the combined nanotube-metal system is then the sum of the RPA irreducible polarizabilities of the two subsystems 
${ P = P^{metal}+P^{CNT}}
\label{Psum}$
which emerges from RPA calculations of the {\it isolated} subsystems \cite{Lischner}.

I define the {\it effective} dielectric function:
\beq
{ \hat{\epsilon}}\equiv1-{ wP^{CNT}}
\label{eps_eff}
\eeq
and the {\it effective} screened Coulomb interaction:
\beq
{ \hat{W}}\equiv{ \hat{\epsilon}}^{-1} { w}
\label{W_w}
\eeq
where 
${ w}=(1-{ VP^{metal}})^{-1}{ V}$
is the screened Coulomb interaction between electrons calculated for the metallic surface alone. 
The effective quantities ${ \hat{W},\hat{\epsilon}^{-1}}$ are by construction equal to the 'true' RPA ones in the nanotube region.

An important feature of the embedding approach is that in order to obtain ${ \hat{\epsilon}}$, ${ \hat{\epsilon}^{-1}}$ or ${ \hat{W}}$ in the nanotube region, one only needs ${ w}$ in the same region, {\it i.e.} outside the metal surface. Here, the 
atomistic properties of the substrate are less important and 
one can use a classical, local theory to obtain ${ w}$. Specifically, one models the metallic surface with a semi-infinite electron gas, characterized by a dielectric function which jumps from unity outside the metal (${ z>z_0}$) to the Drude function ${ \epsilon^D}$ inside (${ z<z_0}$) \cite{Pitarke,Drude}. The location of the "mirror plane" ${ z_0}$ with respect to the position of the atoms of the real metal surface (see inset in Fig. \ref{Fig_17-0_deltaGW}) can be determined separately for a given metal surface. \cite{Needs} Typically ${ z_0}$ is about half the inter-planar spacing above the plane of surface atoms. \cite{Zaremba}

Within this model, ${w}$ takes the following form outside the metal:
\beq
{ w(r,r',\omega)= V(r-r')+g(\omega)/|r-\tilde{r}'|}
\label{w_perf_cond}
\eeq
where ${ g(\omega)\equiv [1-\epsilon^D(\omega)]/[1+\epsilon^D(\omega)]}$ is the surface-response function \cite{Feibelman} and ${ \tilde{r}'}$ is the reflection of ${ r'}$ across the mirror plane. 
Accounting for the frequency dependence of $w$ is critical for the correct description of the renormalization of individual (empty or occupied) QP CNT states; however,  the precise value of the corresponding metal surface plasmon energy is less important for the results presented here.
The use of an {\it ab initio} determined  $z_0$ insures that $w$ is correct  outside the metal to first order in the wavevector $q$ \cite{Liebsch}. Small-$q$ corrections to $g$ enter as ${\mathcal{O}}(z_0q)$ \cite{Feibelman}.  I find that that only wavevectors $|q|\ll z_0^{-1}$ are important for the renormalization effects presented here, suggesting that corrections to $w$ beyond the model can be neglected.

Next I include surface polarization effects into {\it effective} ${ G_0W_0}$/${ BS}$ theories.
The ${ G_0W_0}$ electron self-energy contribution to CNT QP energies  can readily be written as follows:
\beq
{ \langle \Psi_{nk}|{ \Sigma}|\Psi_{mk}\rangle=\langle \Psi_{nk}|{ iG^{CNT}\hat{W}}|\Psi_{mk}\rangle}.
\label{Sigma_eff}
\eeq
where ${ G^{CNT}}$ is the nanotube Green's function and ${ |\Psi_{nk}\rangle}$ the nanotube wavefunction \cite{off-diagonal} for band ${ n}$ and momentum ${ k}$. 
According to Eq. \eqref{Sigma_eff}, ${ G_0W_0}$ calculations of QPs in a nanotube near a metallic surface can be performed as in the isolated CNT case except that ${ V}$ is formally replaced by ${ w}$ when evaluating the nanotube screened Coulomb interaction.

Excitons can also be treated within the embedding approach. It can be shown (see Appendix A) that singlet excitons in the CNT near a metallic surface obey the usual ${ BS}$ equation \cite{Strinati} but with an {\it effective} electron-hole (e-h) interaction kernel ${ \hat{K}}\equiv{ \hat{K}^d+\hat{K}^x}$. The effective direct term  ${ \hat{K}^d}$ can be obtained  by taking the functional derivative of  ${ \Sigma}$ with respect to ${ G^{CNT}}$. Within the usual static approximation its matrix elements between pairs of valence and conduction states read:
\beq
{ \hat{K}^d_{vck,v'c'k'} =  -\langle\Psi^*_{ck}\Psi_{c'k'}| \hat{W}(\omega=0) |\Psi_{vk}\Psi^*_{v'k'}\rangle } 
\label{K_d_reduced}
\eeq
A notable difference from the isolated CNT case is that the effective exchange term ${ \hat{K}^x}$ is  "screened" by ${ w}$:
\beq
{ \hat{K}^x_{vck,v'c'k'} =  2\langle\Psi^*_{ck}\Psi_{vk}| w(\Omega)
|\Psi_{c'k'}\Psi^*_{v'k'}\rangle } 
\label{K_x_reduced}
\eeq
where ${ \Omega}$ is the exciton energy. In this work $\Omega\sim1$ eV. At this energy most metals (including Au) act as near-perfect mirrors and one can replace $w(\Omega)$ by $w(0)$ in Eq. \eqref{K_x_reduced}.

The embedding approach has been implemented in a modified version of the BerkeleyGW package. \cite{BerkeleyGW}
Calculations at large nanotube-surface separations were made possible by a new truncation scheme for ${w}$, described in Appendix C.
The ${ G_0W_0}$ calculations are performed within a mixed technique \cite{Spataru_dop2} in which self-energy effects for the isolated nanotube are calculated within the Generalized Plasmon Pole model (GPP) \cite{Spataru} while changes in ${ \Sigma}$ upon the nanotube approaching the metallic surface are calculated within full frequency RPA. \cite{comp_details}

\section{Results} 
\subsection{Quasiparticle renormalization effects}

\begin{figure}
\resizebox{9.0cm}{!}{\includegraphics[trim=0cm -0.0cm 0cm 0cm,clip=true,angle=0]{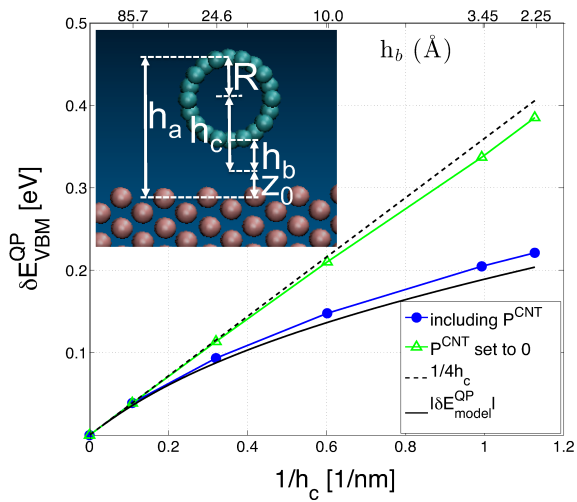}}
\caption{ The change in the QP energy at VBM for the (17,0) CNT, ${ \delta E_{VBM}^{QP} \equiv E_{VBM}^{QP}(h_c)-E_{VBM}^{QP}(\infty)}$, as function of the inverse distance between the center-of-tube and the metal mirror plane.
The inset shows an illustration of a CNT on a metallic surface. The apparent nanotube height ${ h_a}$ is the distance from the top of the CNT to the first metal layer. The distance between the nanotube center and the metal mirror plane is denoted by ${ h_c}$ and $h_b\equiv h_c-R$.}
\label{Fig_17-0_deltaGW}
\end{figure}

Quasiparticles are charged many-electron excitations involving the addition or removal of one electron from the system. When a unit point-charge is brought near/taken away from a perfect conductor from/to infinity, its energy changes by ${ \pm1/(4h)}$ (I use atomic units, {\it i.e.} $e^2/(4\pi\epsilon_0)=1$), where ${ h}$ is the height relative to the metal surface. This classical image-charge picture turns out to describe the QP energy renormalization of molecular orbitals \cite{Neaton,Thygesen1,Thygesen2,Inkson} quite well. As discussed below, a similar model can be extended to CNTs but only after a proper definition of the nanotube-metal separation ${ h}$.

I have studied the QP properties at the valence band maximum (VBM) and conduction band minimum (CBM) of two zig-zag semiconducting CNTs.  I find that the renormalization of the nanotube inverse dielectric function is not important, and the electron self-energy can be well approximated  as ${ \Sigma \approx  iG^{CNT}\epsilon^{-1}_{CNT}w}$, where ${ \epsilon^{-1}_{CNT}}$ is the inverse dielectric function calculated for the isolated CNT. 
Figure \ref{Fig_17-0_deltaGW} shows (circles) the QP energy change ${ \delta E_{VBM}^{QP}}$ at the VBM for the (17,0) CNT (diameter $D=1.32$ nm), calculated \cite{wp_used} as function of the inverse distance ${ h_c}$ between the center of the nanotube and the mirror plane of the metallic surface (see inset in Fig. \ref{Fig_17-0_deltaGW}). The change is significant, reaching more than 200 meV at the smallest separation (${ h_b}=2.25 \ \AA$). A similar renormalization (with opposite sign) is found for the state at CBM. 
Extrapolating the image-charge model to nanotubes, one would expect ${ \delta E_{VBM}^{QP}\approx 1/(4h_c)}$. As indicated by the dashed line in Fig. \ref{Fig_17-0_deltaGW}, this model scaling is appropriate at large separations, but severely overestimates ${ \delta E_{VBM}^{QP}}$ at smaller ones.

One may raise two main issues about the applicability of the image-charge picture to CNTs. First, it is not clear to what extent QPs in a quasi-one dimensional system can be likened to point charges. Second, the image-charge model neglects the screening response of the nanotube. This response is expected to be important in systems with smaller QP bandgap, \cite{Draxl} as they tend to have increased polarizability. 

It is straightforward to disentangle these two effects within our embedding approach. To address the first, I set 
to zero the nanotube polarizability ${ P^{CNT}}$ 
and calculate the resulting self-energy contribution to ${ \delta E^{QP}}$ by evaluating ${ \delta\Sigma^0\equiv {  iG^{CNT}(w-V)}}$ at the QP energy. 
This contribution is indicated by triangles in fig. \ref{Fig_17-0_deltaGW} for the state at VBM, showing good agreement with the image charge model prediction. This can be best understood in terms of Wannier functions. In zig-zag nanotubes these can be maximally localized along the tube axis within the length of the unit cell ${l  \approx 0.4 {\text{ nm}} < D}$ \cite{Marzari}. This implies that it is more appropriate to think of the shape of the added charge as a ring rather than an elongated tubule. For ${ h_c \gg D}$, the electric field due to rings and point charges is the same and in this limit one can show that ${ \delta\Sigma^0_{VBM}\approx 1/(4h_c)}$; at smaller distances, the delocalization of the Wannier functions becomes relevant and yields the rather small deviation from the ${ 1/(4h_c)}$ scaling seen in Fig. \ref{Fig_17-0_deltaGW}.

Clearly, the screening response of the nanotube accounts for most of the difference between the calculated $\delta E^{QP}$ and the ${ \pm1/(4h_c)}$ scaling. To understand this, note that QP energies at CBM/VBM can be written in terms of differences between the ground-state energy of the neutral ($N$ electrons) and charged ($N\pm1$ electrons) system: $E^{QP}=\pm (E^{N\pm1}_0-E^N_0)$. Considering only the dominant interactions ({\it e.g.} neglecting Van der Waals forces) between the nanotube and the surface I estimate $\delta E^{QP}$ within a simple electrostatic model for the term $\delta E^{N\pm1}_0$, {\it i.e.}, from the change in energy of the charged system as one takes it from $\infty$ to near the surface. 

More exactly, one replaces the nanotube with a cylindrical tubule with radius $R=D/2$ and polarizability $\mathsf{\bar{P}}$ that yields the same average (along radial direction) as $P^{CNT}$. \cite{Spataru_dop,Spataru_dop2} The charged system is simulated by adding to the tubule an external unit charge in the shape of a ring with same radius. 
The external charge induces a charge distribution along the tubule, \cite{LinChuu,FL02} assumed for simplicity to be angular symmetric about the tubule axis. Let $h$ be the distance between the tubule axis and the metal mirror plane and $F(h)$ the attractive force between the total charge on the tubule and its mirror image. The model estimates $\delta E^{N\pm1}$ from $\int_{h_c}^{\infty} F(h)\,dh$, leading to the following QP energy renormalization (see Appendix B):
\beq
\delta E^{QP}_{\text{model}}=\pm \frac{1}{2}  \int \frac{dq}{2\pi} \,\frac{\mathsf{\bar{w}}(q)-\mathsf{\bar{V}}(q)}{\mathsf{\bar{\epsilon}}(q)\mathsf{\bar{\epsilon}_0}(q)}
\label{electrostatic_model}
\eeq
where $\mathsf{\bar{\epsilon}}=1-\mathsf{\bar{P}}\mathsf{\bar{w}}$ and $\mathsf{\bar{\epsilon}}_0=1-\mathsf{\bar{P}}\mathsf{\bar{V}}$ with $\mathsf{\bar{V}}(q)=2I_0(Rq)K_0(Rq)$ and $\mathsf{\bar{w}}(q)-\bar{V}(q)=-2K_0(2h_cq)$ ($I_0$ and $K_0$ are the modified Bessel functions of the first and second kind). The exponential decay of the integrand 
for $|q|>h_c^{-1}$ implies that it is sufficient to consider only the small $q$ behavior ($|q|\ll2\pi/l$) of $\mathsf{\bar{P}}$, {\it i.e.} one can set $\mathsf{\bar{P}}=\alpha q^2$, where $\alpha$ is the static polarizability of the nanotube \cite{Benedict_alpha}. 
One uses $\alpha= a_0+a_1 R^2$ with $a_0=38.0\, \AA^2$ and $a_1=6.92$, as suggested by previous {\it ab initio} studies of a large variety of semiconducting CNTs. \cite{Guo} Applied to the (17,0) CNT, the electrostatic model yields a QP renormalization in very good agreement with the {\it ab initio} results, as shown in  Fig.\ref{Fig_17-0_deltaGW}. 
 
\begin{figure}
\resizebox{8.0cm}{!}{\includegraphics[trim=0cm -0cm 0cm 0cm, clip=true,angle=0]{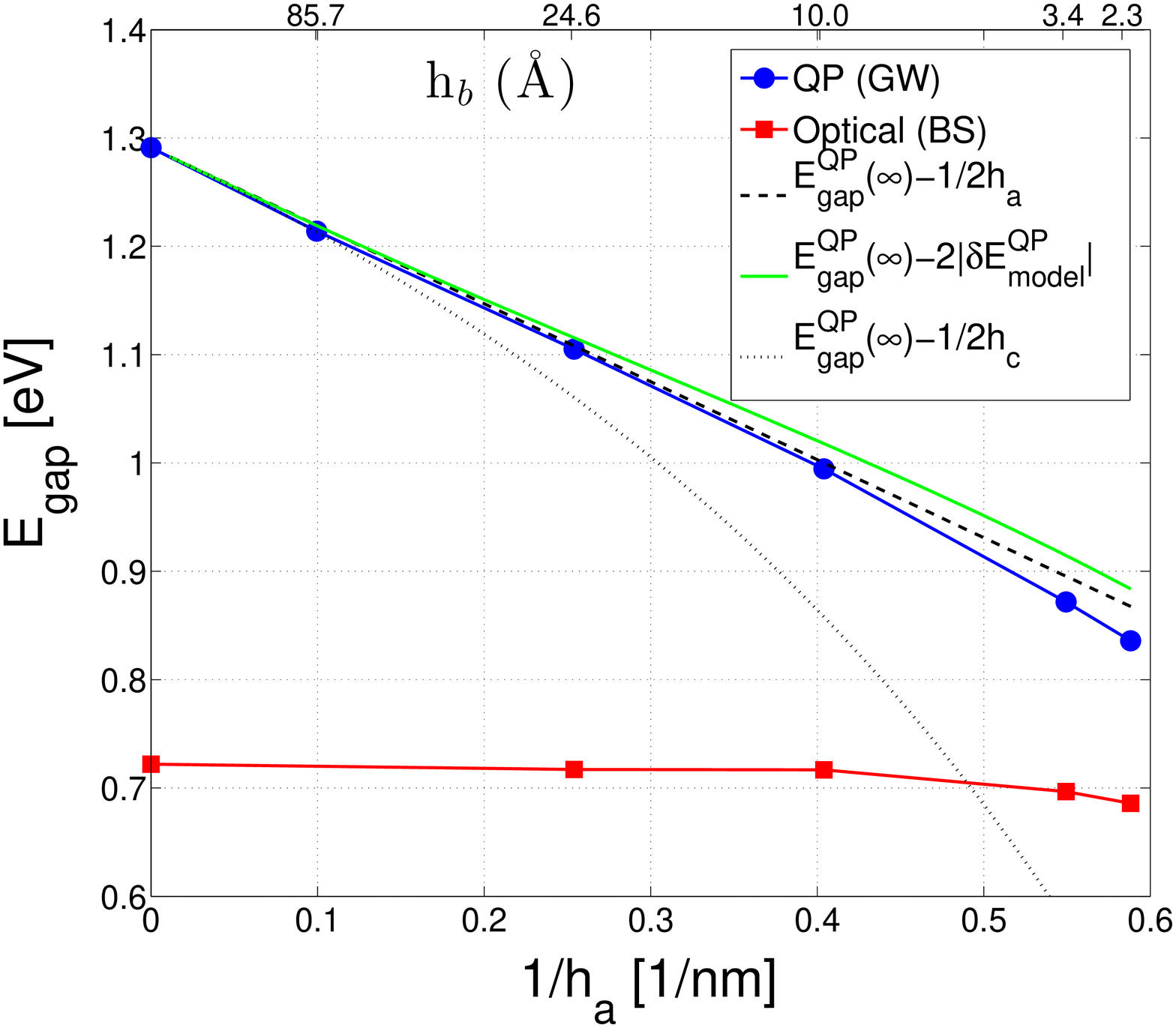}}
\resizebox{8.0cm}{!}{\includegraphics[trim=0cm -0cm 0cm 0cm, clip=true,angle=0]{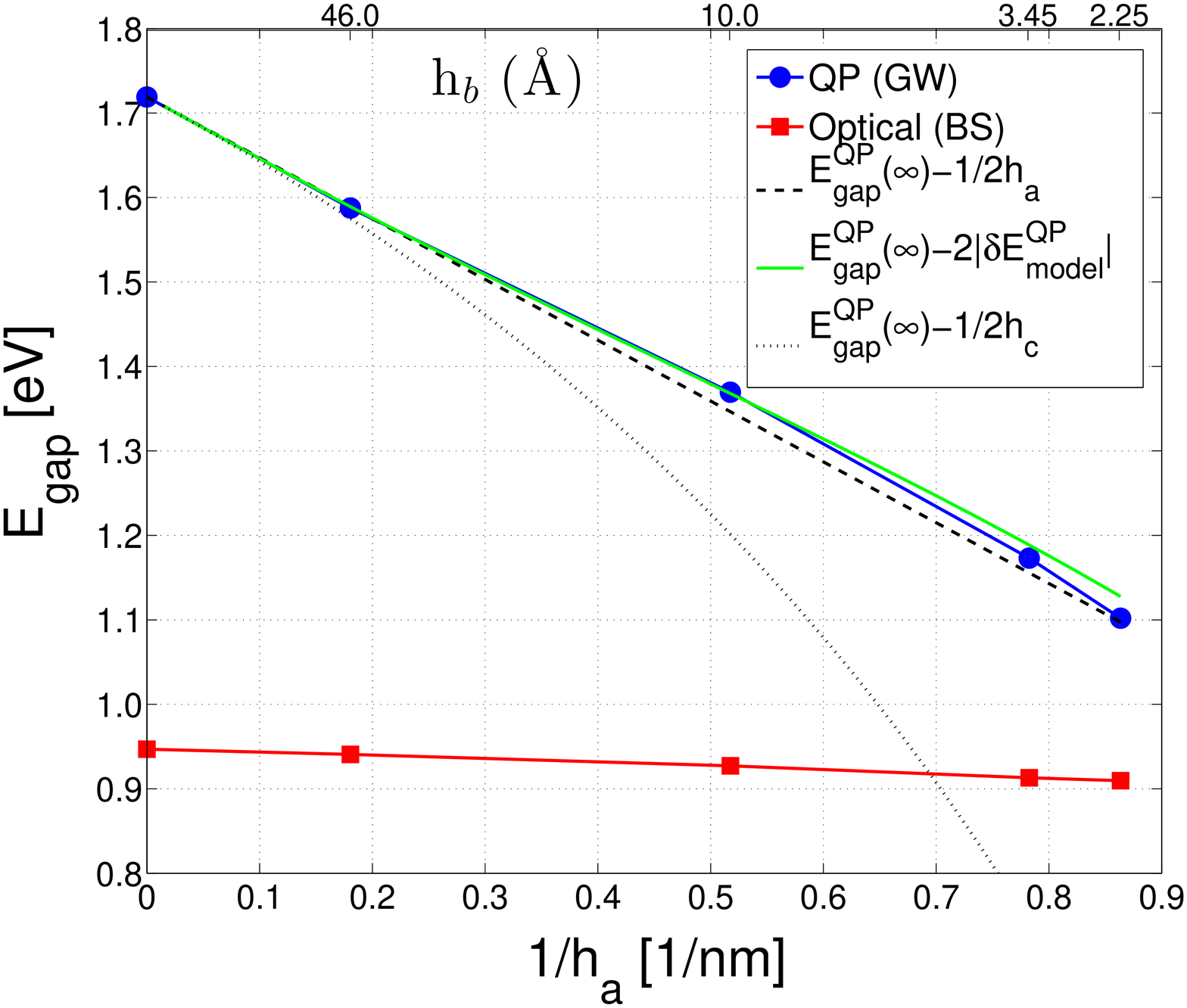}}
\caption{QP and optical bandgaps of the (17,0) CNT (a) and (10,0) CNT (b) near a metallic surface as function of the inverse apparent nanotube height ${ h_a}$ (${ z_0=1.5 \AA}$ as appropriate for a Au surface). Exciton binding energies are equal to the energy differences between the QP bandgap (circles) and the optical gap (squares).
}
\label{Fig_17-0_10-0_ha}
\end{figure}

QP bandgap renormalization (BGR) in semiconducting CNTs has been recently studied via scanning tunneling spectroscopy of a nanotube bundle on Au(111) \cite{Lin}. The QP bandgap was found to vary inversely with the apparent nanotube height, with the bundle playing the role of a spacer. Inspired by these results, note in Fig. \ref{Fig_17-0_10-0_ha}(a) the calculated QP bandgap of the of the (17,0) CNT as a function of the apparent nanotube height ${ h_a}$, defined as the distance between the top of the nanotube and the first layer of the metal surface (alternatively, one can think of ${ h_a}$ as the distance between the top of a $\pi$ orbital at the top of the tube and the metal mirror plane). I choose ${ z_0=1.5 \AA}$ as appropriate for Au(111). \cite{Tamblyn} Comparison with the dashed curve shows  that the change in QP bandgap ${ \delta E_{gap}^{QP}}$ is well described by the ${ -1/(2h_a)}$ scaling [as opposed to the ${ -1/(2h_c)}$ scaling, which does not reflect the nanotube screening response; see the dotted curve]. This suggests that the image-charge model can be extended to CNTs by incorporating the effect of the nanotube screening response into an appropriate definition of the CNT-surface distance, namely ${ h\equiv h_a}$. 

To check the applicability of the ${ -1/(2h_a)}$ scaling to QP BGR in other CNTs, I have also studied the smaller diameter (10,0) CNT (${ D=}0.78$ nm). As shown in Fig. \ref{Fig_17-0_10-0_ha}(b), the QP bandgap changes significantly by more than $0.6$ eV from ${ 1.72}$ eV in the isolated case to ${ 1.10}$ eV at weak physisorption distances.
The scaling ${ \delta E_{gap}^{QP} \approx -1/(2h_a)}$ holds for this tube as well, suggesting its 'universality' across semiconducting nanotubes with different diameters. To test this assumption, I use Eq. \eqref{electrostatic_model} to estimate QP BGR for a wide range of CNT diameters. The agreement between the model and the  $-1/(2h_a)$ scaling is excellent for the (17,0) and (10,0) CNTs,  as shown in Fig. \ref{Fig_17-0_10-0_ha}. In general, for any practical nanotube diameter I find that the agreement is very good (with the largest difference equal to $\sim10\%$ in the limit $R\rightarrow\infty$) (see Appendix B), in strong support of the above assumption.

The (17,0) CNT has a similar diameter to those from the experiment in Ref. \cite{Lin}. Its calculated ${ E_{gap}^{QP}}$ ranges from $1.29$ (isolated tube) to $0.83$ eV (physisorbed tube). This can be compared with the experimental values, namely ${ 1.1}$ eV (value extrapolated for nanotubes far away from the surface) and $0.73$ eV (for nanotubes in contact to the surface). The difference between theory and experiment is explained by additional environmental effects such as screening from other nanotubes (including metallic ones). \cite{Lin}

\subsection{Exciton renormalization effects}

Next I consider the renormalization of excitons in nanotubes upon physisorption on a metallic surface. I consider the lowest energy singlet excitations induced by light polarized along the nanotube axis. 
The attractive, direct term ${ \hat{K}^D}$ dominates the e-h inteaction and is affected by the surface via the change in the static effective screened Coulomb interaction ${ \delta \hat{W}\approx \epsilon^{-1}_{CNT}(w-V)}$. As the nanotube approaches the surface, the coupling between e-h pairs  decreases due to their induced image in the metal surface and the exciton gets more delocalized while its binding energy ${ E_b}$ diminishes accordingly. 

\begin{figure}
\resizebox{8.0cm}{!}{\includegraphics[trim=0cm -0cm 0cm 0cm, clip=true,angle=0]{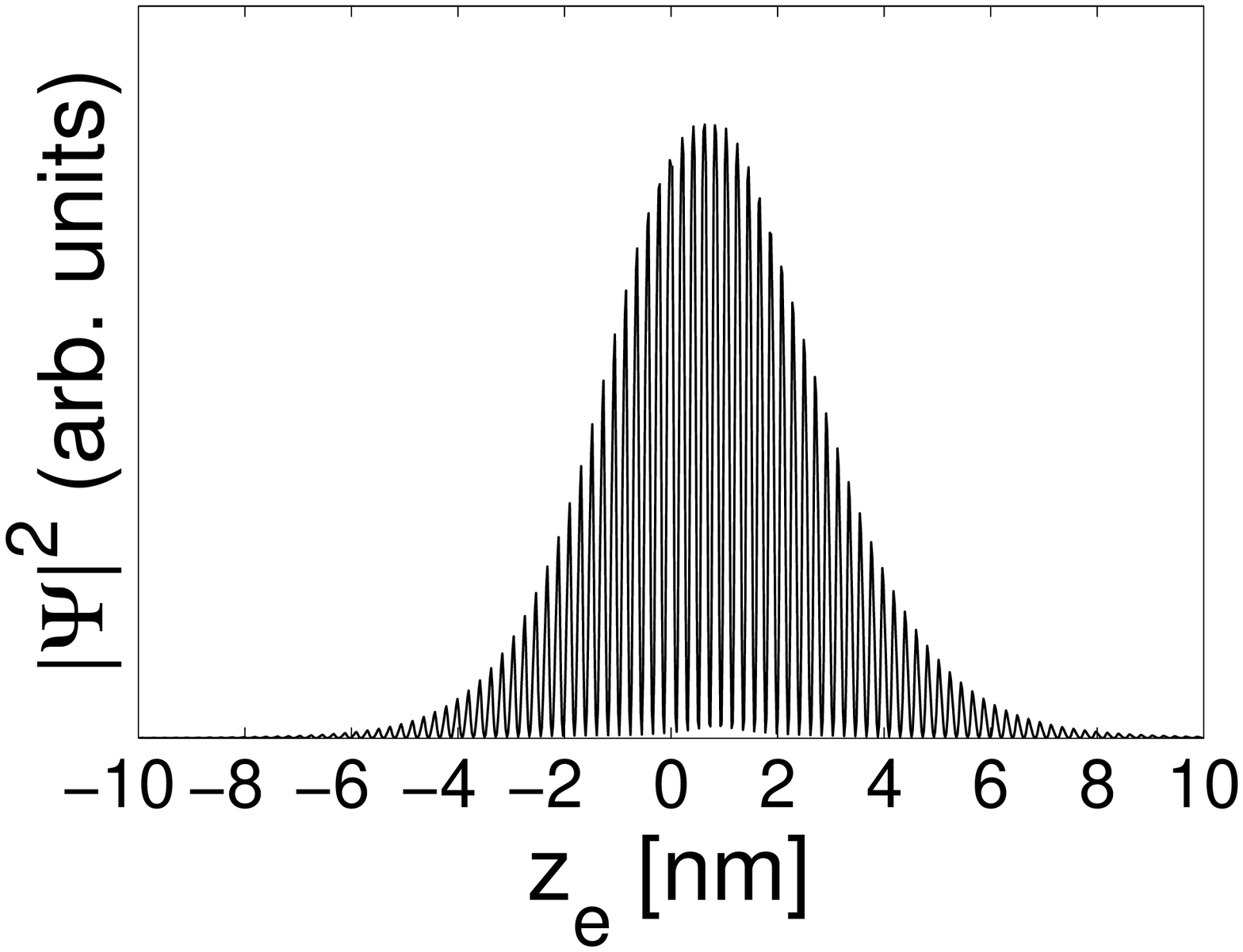}}
\resizebox{8.0cm}{!}{\includegraphics[trim=0cm -0cm 0cm 0cm, clip=true,angle=0]{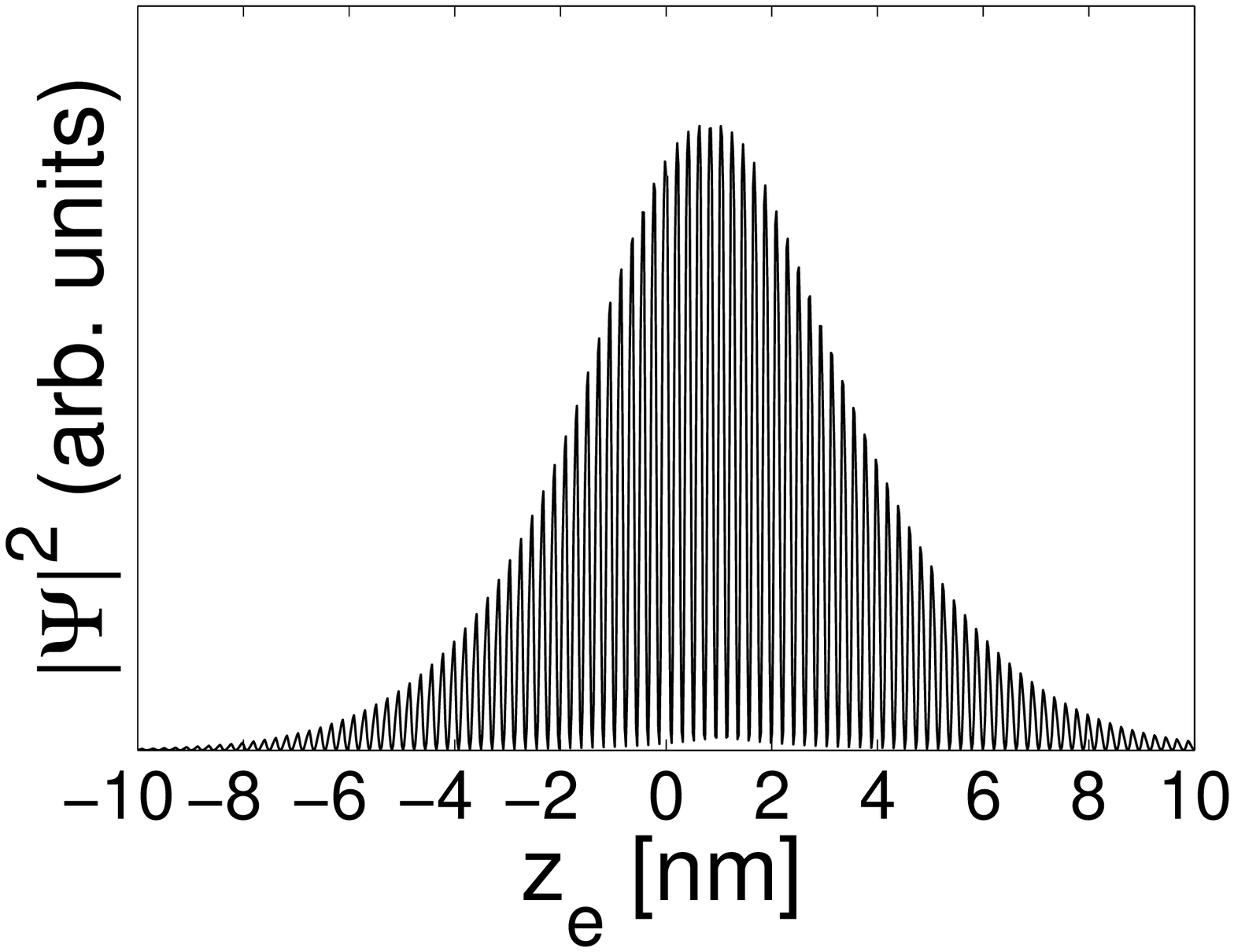}}
\caption{Exciton wavefunction of the brightest lowest energy exciton in the (17,0) CNT far away from (a) and physisorbed on (b) a metal surface. 
}
\label{Fig_17-0_excitons}
\end{figure}

Figure \ref{Fig_17-0_excitons} shows the exciton wavefunction of the brightest lowest energy exciton in the (17,0) CNT, far away from [Fig.\ref{Fig_17-0_excitons}(a)] and physisorbed on [Fig.\ref{Fig_17-0_excitons}(b)] a metal surface. The plots show the probability of finding an electron at a distance $z_e$ away from a hole, obtained by fixing the hole near a carbon atom (the plots are insensitive to the location of the carbon atom relative to the metal surface) after integrating out the electron coordinates along directions perpendicular to the tube axis. The size of the exciton (the root mean square of exciton envelope function) changes by $\sim35\%$, from $\sim2$ nm for the isolated case to $\sim2.7$ nm in the physisorbed case. 

For weak physisorption, excitons still bind but their binding energy  is about $75\%$ smaller than in the isolated case: ${ E_b}$ changes in the (17,0) CNT from ${0.57}$ eV (isolated tube) to ${0.15}$ eV (physisorbed tube), while for the (10,0) CNT these values are ${0.77}$ and ${0.19}$ eV.

Compensation between self-energy and excitonic effects \cite{compensation_refs}  results in order of magnitude smaller renormalization of the nanotube optical gap compared to QP BGR, as indicated by red squares in Fig. \ref{Fig_17-0_10-0_ha}(a) and \ref{Fig_17-0_10-0_ha}(b). This is because the excitons are neutral with no overall static dipole. \cite{Thygesen1} The slight inhomogeneous charge distribution of the exciton \cite{Rohlfing_2tubes} gives rise to a small (less than $50$ meV) red-shift in the optical gap for both  (10,0) and (17,0) CNTs. Similarly, I find that the metal surface induces negligible renormalization of the local transition dipole matrix element \cite{Spataru_rad} of the brightest lowest energy CNT exciton.

\section{Summary}
In conclusion, I have developed an embedding approach that allows ${ G_0W_0}$ or ${ BS}$ calculations to reach beyond the computational limits of standard computations, making possible the study of electronic and optical properties of nanostructures near a metallic surface with relatively small computational cost. Applied to the case of semiconducting CNTs it shows that a metallic surface can induce strong renormalization of QP and exciton binding energies. The renormalization scales inverse proportionally to the nanotube apparent height, with the nanotube screening response playing an important role in establishing the effect. This addresses fundamental aspects of the physics of the CNT/metal contact and should contribute toward a better understanding of CNT-based devices. One possible application is the realization of a heterojunction superlattice within a nanotube by deposition on a surface patterned with different dielectric properties.

\begin{acknowledgments}
I thank P. Feibelman, F. L\'{e}onard, M. Shaughnessy and A. Talin for a critical reading of the manuscript and/or useful discussions.
Work supported by the Early Career Laboratory Directed Research and Development program at Sandia National Laboratories, a multiprogram laboratory managed and operated by Sandia Corporation, a wholly owned subsidiary of Lockheed Martin Corporation, for the United States Department of Energy's National Nuclear Security Administration under Contract No. DE-AC04-94AL85000.
\end{acknowledgments}

\renewcommand{\theequation}{A-\arabic{equation}} 
\section*{Appendix A: Effective Bethe-Salpeter equation for excitons}
\setcounter{equation}{0}

Here I derive the Bethe-Salpeter equation (BSE) for excitons in the CNT near a metallic surface. Let us start from the BSE for the two-particle correlation function $L$ (generalized coordinates denote space, time, and spin variables), \cite{Strinati}:
\begin{multline}
{ L (1,2;1',2') =  G(1,2)G(1',2') + }
\\
{ \int {\rm d}
(3456) G(1,4)G(1',3) { K} (3,5;4,6) L (6,2;5,2')}
\label{BS_L_complete}
\end{multline}
with ${ K}$ the electron-hole (e-h) interaction kernel:
\beq
{ K(1,2;3,4) \equiv  { {\rm \delta} [ V_H(1,3) +\Sigma(1,3) ]
\over {\rm \delta} G(4,2) }}
\label{K_funct_der_complete}
\eeq
where ${ V_H}$ is the Hartree potential. Within the ${ GW}$ approximation for ${ \Sigma}$ and to first-order perturbation expansion about ${ W}$, ${ K}$ is the sum between the usual exchange  and direct terms \cite{Strinati}: 
\begin{multline}
{ K (1,2;3,4) = -i \delta(1,3) \delta(2,4)  V(1,2) +} 
\\
{i \delta(1,4) \delta(3,2) W(1^+,2)}
\label{K_complete}
\end{multline}

One is interested in renormalization effects induced by the metallic surface on excitons in the CNT. It is possible to derive in the reduced CNT space region an effective equation for the polarizability ${ \chi(1,2) \equiv = -i L (1,2;1^+,2^+)}$ with ${ r_1,r_2} \in \text{CNT}$. This can be done in two steps as follows.

In the first step I write the BSE for the {\it irreducible} two-particle correlation function ${ \tilde{L}}$, which obeys \cite{Strinati} a similar equation as \eqref{BS_L_complete} except that ${ K}$ replaced by ${ K^d}$.  The BSE for ${ \tilde{L}}$ then can be written in the reduced space of the CNT region \cite{non-cross} as
\begin{multline}
{ \tilde{L} (1,2;1',2') =  G^{CNT}(1,2)G^{CNT}(1',2') + }
\\
{ i\int {\rm d}
(34) G^{CNT}(1,4)G^{CNT}(1',3) {\hat W} (3^+,4) \tilde{L} (3,2;4,2')}
\label{BS_irredL}
\end{multline}
The solution of eq. \eqref{BS_irredL} yields the CNT irreducible polarizability: ${ \tilde{\chi}(1,2) \equiv -i \tilde{L} (1,2;1^+,2^+)}$.

In the second step I write the equation for the full polarizability, $\chi(1,2) $\cite{Strinati}
\beq
{ \chi(1,2)=\tilde{\chi}(1,2)+\int {\rm d}(34) \tilde{\chi}(1,3)V(3,4)\chi(4,2)}
\label{ful_chi}
\eeq
Note that the space-integral in Eq. \eqref{ful_chi} involving the ${ r_4}$ coordinate runs over both the CNT and the metal regions. To arrive at an equation in the reduced CNT space, one needs to integrate out the ${ r_4}$ coordinate over the metal region. This is straightforward to achieve after safely assuming that the irreducible polarizability in the metal surface region is not renormalized appreciably by the presence of the CNT, {\it i.e.}, ${ \tilde{\chi}(1,2) =\tilde{\chi}^{metal}(1,2)}$ for ${ r_1,r_2\in\text{metal}}$ where ${ \tilde{\chi}^{metal}}$ is the irreducible polarizability of the isolated metal surface. For
${ r_4 \in\text{metal}}$ and ${r_2 \in\text{CNT}}$ 
one then can write
\beq
{ \chi(4,2)=\int {\rm d}(2'4') {\chi}^{metal}(4,4')V(4',2')\chi(2',2)}
\label{ful_chi2}
\eeq
where the $r_{2'}$ coordinate runs over the CNT region and
$\chi^{metal}$ (the full polarizability of the isolated metal surface) is defined by:
\begin{multline}
 \chi^{metal}(1,2)=\tilde{\chi}^{metal}(1,2)+
\\
\int {\rm d}(34) \tilde{\chi}^{metal}(1,3)V(3,4)\chi^{metal}(4,2)
\label{ful_chi_metal}
\end{multline}
Combining \eqref{ful_chi} and \eqref{ful_chi2} and
using that $w\equiv V+V\chi^{metal}V$ 
one arrives at the following effective equation for $\chi$ in the CNT region:
\beq
{ \chi(1,2)=\tilde{\chi}(1,2)+\int {\rm d}(34) \tilde{\chi}(1,3)w(3,4)\chi(4,2)}
\label{red_chi}
\eeq
where all the spatial coordinates in Eq. \eqref{red_chi} are in the CNT region. 

The set of Eqs. \eqref{BS_irredL}-\eqref{red_chi} are equivalent to the following BSE with all coordinates in the reduced CNT space:
\begin{multline}
{ L (1,2;1^+,2^+) =  G^{CNT}(1,2)G^{CNT}(1^+,2^+) + }
\\
{ \int {\rm d}
(3456) G^{CNT}(1,4)G^{CNT}(1',3) { \hat{K}} (3,5;4,6) L (6,2;5,2')}
\label{BS_L_reduced}
\end{multline}
where the {\it effective} e-h interaction kernel ${ \hat{K}}\equiv { \hat{K}^x+\hat{K}^d}$ is:
\beq
{ \hat{K}} (1,2;3,4) = -i \delta(1,3) \delta(2,4)  w(1,2) + i \delta(1,4) \delta(3,2) \hat{W}(1^+,2)
\label{K_reduced}
\eeq

The reduced BSE Eq. \eqref{BS_L_reduced} is solved within the usual Tamm-Damcoff approximation.
The result for singlet excitons in the CNT near the metallic surface is an effective BSE with an effective direct term written in Fourier space (within the usual static approximation, valid when the binding energy of excitons is significantly smaller than the energy of plasmons),
\beq
{ \hat{K}^d ( r_1,r_2;r_3,r_4) =  i \delta(r_1,r_4) \delta(r_3,r_2) \hat{W}(r_1,r_2,\omega=0)}  
\label{K_d_reduced_1}
\eeq
and an effective exchange term "screened" \cite{Benedict} by the metal surface,
\beq
{ \hat{K}^x (r_1,r_2;r_3,r_4) =  -2i \delta(r_1,r_3) \delta(r_2,r_4)  w(r_1,r_2,\Omega)} 
\label{K_x_reduced_1}
\eeq
where ${ \Omega}$ is the energy of the exciton. For  ${ \Omega}$ significantly smaller than 
the metal surface plasmon energy, one can further replace $w(\Omega)$ by $w(0)$ in Eq. \eqref{K_x_reduced_1}.
 
The effective BSE \eqref{K_d_reduced_1} and \eqref{K_x_reduced_1} can be used for the study of excitons in other nanostructures near a metal surface, as long as the relevant electronic states of the nanostructure do not overlap significantly with those of the metal surface.

\renewcommand{\theequation}{B-\arabic{equation}} 
\section*{Appendix B: Electrostatic model for QP renormalization}
\setcounter{equation}{0}

Consider the case where an external ring-shaped unit charge is added on a polarizable tubule with tube axis situated at a distance $h$ away from the mirror plane of a metal surface. 
The electrostatic model assumes angular symmetry about the tubule axis. 
The added unit charge ring induces on the tubule a charge density:
\beq
n^{ind}(q)=\mathsf{\bar{P}}(q) V_{tot}(q)
\eeq
 where $\mathsf{\bar{P}}$ is the irreducible polarizability of the tubule and $V_{tot}$ is the total potential produced by the charged ring in the presence of the metal surface. The induced charge shows sign oscillations along the tube axis and for a semiconducting nanotube it integrates to zero \cite{metallicCNT}. 

The total potential is given by
\beq
V_{tot}(q)=\bar{w}(q)/\bar{\epsilon}(q)
\eeq
where $\bar{\epsilon}=1-\mathsf{\bar{P}}\bar{w}$ and $\bar{w}$ is the Coulomb interaction produced by the added external ring and its polarization on the metal surface.  Let us denote by $\mathsf{\bar{V}}$ the Coulomb interaction between two unit charge rings  on the tubule; its Fourier transform  has the usual form: $\mathsf{\bar{V}}(q)=2I_0(qR)K_0(qR)$. The remaining interaction ($\bar{w}-\mathsf{\bar{V}}$) between a ring and the image of the other ring can be approximated very well by the interaction between two point charges, one at the center of one ring and another at the center of the image of the other ring [for the (17,0) CNT, this approximation affects the QP renormalization results by $\sim1\%$]. The result is:
\beq
\bar{w}(q)-\mathsf{\bar{V}}(q) \equiv -\int \, dz \frac{e^{-iqz}}{\sqrt{z^2+(2h)^2}}=-2K_0(2hq)
\eeq

The polarization per unit tube length $\mathsf{\bar{P}}$ can be obtained from the $G=G'=0$ components of the {\it ab initio} calculated $P^{CNT}_{GG'}(q)$ as in Ref. \cite{Spataru_dop}: $\mathsf{\bar{P}}(q)=P^{CNT}_{00}(q) A_{uc}$, where $A_{uc}$ is the cross-sectional area of the unit cell used in the {\it ab initio} calculation.

The attractive force $F(h)$ between the charged tubule and its mirror image can be written
\beq
F(h)=\int \,\frac{dq}{2\pi} \lambda(q)^2 \frac{d}{d(2h)}[\bar{w}(q)-\mathsf{\bar{V}}(q)]
\eeq
where the total charge distribution on the tubule is $\lambda\equiv 1+n^{ind}=1/\bar{\epsilon}$.
Integrating the force from $h_c$ to $\infty$ leads to
\beq
\delta E^{QP}_{model}
=\pm\frac{1}{2}\int \,\frac{dq}{2\pi} \int_{h_c}^{\infty} dh \, \frac{1}{\bar{\epsilon}(q)^2} \frac{d}{dh}\bar{w}(q)
\eeq

The integral over $h$ can be written as ($\mathsf{\bar{w}}\equiv\bar{w}{\vline_{h=h_c}}$)
\beq
\int_{\mathsf{\bar{w}}}^\mathsf{\bar{V}} d \bar{w} \frac{1}{(1-\mathsf{\bar{P}}\bar{w})^2}=\frac{\mathsf{\bar{w}} -\mathsf{\bar{V}} }{(1-\mathsf{\bar{P}}\mathsf{\bar{w}})(1-\mathsf{\bar{P}}\mathsf{\bar{V}})}
\eeq
and one finally obtains
\beq
\delta E^{QP}_{model}=\pm\frac{1}{2}\int \,\frac{dq}{2\pi} \frac{\mathsf{\bar{w}}(q) -\mathsf{\bar{V}(q)} }
{[1-\mathsf{\bar{P}}(q)\mathsf{\bar{w}}(q)][1-\mathsf{\bar{P}}(q)\mathsf{\bar{V}}(q)]
}
\eeq
which is the same as Eq. (7) of the main text. 

The fast decay of the integrand [$K_0(2h_cq) \sim \exp(-2h_c|q|)$] 
for $|q|>h_c^{-1}$ implies that it is sufficient to consider only the small $q$ behavior ($|q|\ll2\pi/l$) of $\mathsf{\bar{P}}$, {\it i.e.} one can set $\mathsf{\bar{P}}=\alpha q^2$ where $\alpha$ is the static polarizability of the nanotube \cite{Benedict_alpha} for electric field parallel to the tube axis. 
I use $\alpha= a_0+a_1 R^2$ with $a_0=38.0\, \AA^2$ and $a_1=6.92$ independent of tube chirality, as suggested by previous {\it ab initio} studies of a large variety of semiconducting CNTs. \cite{Guo} 

I find numerically that within less than $10 \%$, $\delta E^{QP}_{model} \approx \pm1/(4h_a)$  for any practical $R>0.25$ nm. In particular, for large diameter CNTs the model predicts $\delta E^{QP}_{model} \cong \pm \frac{0.94}{4h_a}$ for a nanotube near the surface ($h_c\cong R$). \cite{largeR}

\renewcommand{\theequation}{C-\arabic{equation}} 
\section*{Appendix C: Truncation scheme for the screened Coulomb interaction of the metal surface $w$}
\setcounter{equation}{0}

The bare Coulomb interaction $V$ has been truncated around the nanotube as usual \cite{Spataru_ApplPhysA,Sohrab} and 
I have used the following rectangular truncation scheme for $w-V$ (the nanotube axis is parallel to the {\it x}axis and the metal surface is perpendicular to the {\it z} axis; $z$ coordinates are measured from the mirror plane):
\beq
\frac{1}{|r-\tilde{r}'|} \rightarrow \frac{f_1(x,x')f_2(y,y')f_3(z,\tilde{z}')}{\sqrt{(x-x')^2+(y-y')^2+(z-\tilde{z}')^2}}
\eeq
with:
\beq
f_1(x,x')=\theta(N_kL_1/2-|x-x'|)\theta(L_1-|x+x'|) \\
\nonumber
\eeq
\beq
f_2(y,y')=\theta(L_2/2-|y-y'|)\theta(L_2-|y+y'|)
\nonumber
\eeq
\beq
f_3(z,\tilde{z}')=\theta(L_3-|z-\tilde{z}'|)\theta(z-\tilde{z}')\theta(L_3/2-|z+\tilde{z}'|)
\eeq
where $\theta$ is the step function, $N_k$ is the number of $k$ points used to sample the one-dimensional Brillouin zone, and $L_1$, $L_2$, and $L_3$ are the dimensions of the unit cell along the {\it x},{\it y}, and {\it z} axes respectively. One assumes that the nanotube region is situated between $z=0$ and $z=L_3/2$.  I have used $L_1=l=0.42$ nm, $L_2=1.8\backslash 2.8$ nm and $L_3=2.5\backslash 3.7$ nm for the (10,0)$\backslash$(17,0) CNT. A change of variables has been applied at large nanotube-surface separations $h_b$,  shifting the $z$, $z'$ coordinates by an appropriate distance $d\lesssim h_b$.

The above truncation scheme reduces the computation of the six-dimensional (6D) Fourier transform to the numerical evaluation of three real-space integrals (the other three are analytical) and requires a relatively large $L_3>L_2$. An alternative scheme is to truncate $1/|r-\tilde{r}'|$ separately along each of the $z$, $z'$ coordinates; this scheme requires a significantly smaller $L_3$ but four real-space integrals need to be numerically evaluated for the 6D Fourier transform.


\begin{thebibliography}{99}
\bibitem{CNTbook}
M. Endo, M.S. Strano, and P.M. Ajayan, {Topics  Appl. Physics} {\bf 111} , 13 (2008),  in {\it Carbon Nanotubes} edited by A. Jorio, G. Dresselhaus, M. S. Dresselhaus (Springer, Berlin, 2008), and references therein.
\bibitem{Spataru_dop}
C. D. Spataru and F. L\'{e}onard, {Phys. Rev. Lett.} {\bf 104%
}, 177402 (2010).
\bibitem{Rohlfing_2tubes}
M. Rohlfing, {Phys. Rev. Lett.} {\bf 108}, 087402 (2012).
\bibitem{Ando_env}
T. Ando, J. Phys. Soc. Jpn. {\bf 79}, 024706 (2010).
\bibitem{SpataruFLstrain}
C.D. Spataru and F. L\'{e}onard,
{Phys. Rev. B} {\bf 88}, 045404 (2013).
\bibitem{Leonard}
F. L\'{e}onard and A.A. Talin, Nat. Nanotechnology, {\bf 6}, 773 (2011).
\bibitem{Lin}
H. Lin, J. Lagoute, V. Repain, C. Chacon, Y. Girard, J.-S. Lauret, F. Ducastelle, A. Loiseau and S. Rousset, {Nat. Mater.} \textbf{9}, 235 (2010).
\bibitem{Wang}
F. Wang, G. Dukovic, L.E. Brus and T.F. Heinz, {Science}. \textbf{308}, 838 (2005).
\bibitem{Zojer}
G. Heimel, L. Romaner, J.-L. Bredas, and E. Zojer, {Phys. Rev. Lett.} {\bf 96}, 196806 (2006);
\bibitem{Hong}
G. Hong, S.M. Tabakman, K. Welsher, H. Wang, X. Wang and H. Dai, {J. Am. Chem. Soc.} \textbf{132}, 15920 (2010).
\bibitem{Sakashita}
T. Sakashita, Y. Miyauchi, K. Matsuda and Y. Kanemitsu, {Appl. Phys. Lett.} \textbf{97}, 063110 (2010).
\bibitem{Clair}
S. Clair, Y. Kim and M. Kawai, Phys. Rev. B {\bf 83}, 245422 (2011).
\bibitem{Cummings}
A.W. Cummings and F. L\'{e}onard, ACS Nano {\bf 6}, 4494, 2012).
\bibitem{Spataru_dop2}
C. D. Spataru and F. L\'{e}onard, Chem. Phys. \textbf{413}, 81 (2013).
\bibitem{Spataru}
C.D. Spataru, S. Ismail-Beigi, L.X. Benedict, S.G. Louie, {Phys. Rev. Lett.} \textbf{92},  077402 (2004).
\bibitem{Ando} 
T. Ando, J. Phys. Soc. Jpn. \textbf{66}, 1066 (1997).
\bibitem{Chang}
E. Chang, G. Bussi, A. Ruini, E. Molinari, Phys. Rev. Lett. {\bf 92}, 196401 (2004).
\bibitem{Perebeinos}
V. Perebeinos, J. Tersoff, P. Avouris, Phys. Rev. Lett. {\bf 92}, 257402 (2004).
\bibitem{Zhao} 
H. Zhao, S. Mazumdar, Phys. Rev. Lett. {\bf 93}, 157402 (2004).
\bibitem{HL}
M.S. Hybertsen and S.G. Louie, {Phys. Rev. B} \textbf{34}, 5390 (1986).
\bibitem{Rohlfing}
M. Rohlfing and S.G. Louie, {Phys. Rev. B} \textbf{62}, 4927 (2000).
\bibitem{KS}
W. Kohn, L.J. Sham, {Phys. Rev.} 140 (1965) 1133.
\bibitem{QEspresso} P. Giannozzi et al., J.Phys.:Condens.Matter 21, 395502 (2009).
\bibitem{Lischner}
J. Lischner, D. Vigil-Fowler and S.G. Louie, Phys. Rev. Lett. {\bf 110}, 146801 (2013).
\bibitem{Pitarke}
J.M. Pitarke, V.M. Silkin, E.V. Chulkov, and P.M. Echenique, J. Opt. A: Pure Appl. Opt. {\bf 7}, S73-S84 (2005); J.M. Pitarke et al., {Phys. Rev. B} {\bf 70}, 205403 (2004).
\bibitem{Drude}
The Drude dielectric function ${ \epsilon^D(\omega)=1-\omega_p^2/[\omega(\omega+i\eta)]}$ depends on the bulk plasma frequency of the metal ${ \omega_p}$. I find that QP BGR in nanotubes is insensitive (within 10 meV) to ${ \omega_p}$ in the range considered in this work, namely from $8.55$ eV (Au) to $15.3$ eV (Al).
\bibitem{Needs}
S.C. Lam and R.J. Needs, J. Phys. Condens. Matter \textbf{5}, 2101 (1993).
\bibitem{Zaremba}
E. Zaremba and W. Kohn, {Phys. Rev. B} \textbf{13}, 2270 (1976).
\bibitem{Feibelman} 
P.J. Feibelman, Progr. Surf. Sci. 12, 287 (1982). {\it ibid.} Phys. Rev. B {\bf 22} 3654 (1980).
\bibitem{Liebsch}
A. Liebsch, {\it Electronic Excitations at Metal Surfaces} (Plenum, New York, 1997)
\bibitem{off-diagonal}
I have checked that near the bandgap the nanotube LDA wavefunctions approximate very well the QP ones for both isolated and physisorbed case.  
\bibitem{Strinati}
G. Strinati, Rivista Del Nuovo Cimento \textbf{11}, 1 (1988).
\bibitem{BerkeleyGW} J. Deslippe, G. Samsonidze, D.A. Strubbe, M. Jain, M.L. Cohen, S.G. Louie, Comp. Phys. Comm. {\bf 183}, 12692012 (2012).
\bibitem{comp_details}
The one-dimensional Brillouin zone has been sampled by $32$ ${ k}$ points in the full-frequency RPA calculations,by up to $128$ ${ k}$ points in the GPP ones and by up to $512$ ${ k}$ points when solving the ${ BS}$ equation. The frequency-dependent dielectric matrix has been calculated using up to $300$ empty bands and a $3$ Ry energy cutoff 
and has been sampled along the real-frequency axis using a fine grid with a spacing of $0.1$ eV up to $10$ eV and a coarser grid up to $40$ eV. I have included four valence and four conduction bands in the $BS$ calculations. 
\bibitem{Neaton}
J.B. Neaton, M.S. Hybertsen, and S.G. Louie, {Phys. Rev. Lett.} {\bf 97}, 216405 (2006).
\bibitem{Thygesen1}
J.M. Garcia-Lastra and K.S. Thygesen, {Phys. Rev. Lett.} {\bf 106}, 187402 (2011).
\bibitem{Thygesen2}
J.M. Garcia-Lastra, C. Rostgaard, A. Rubio, and K.S. Thygesen, {Phys. Rev. B} {\bf 80}, 245427 (2009).
\bibitem{Inkson} J.C. Inkson, J. Phys. C {\bf 6}, 1350 (1973).
\bibitem{wp_used}
The results in Fig. \ref{Fig_17-0_deltaGW} are obtained using ${ \omega_p}=8.55$ eV; they depend weakly (within $20$ meV) on the metal type as one changes ${ \omega_p}$ from $8.55$ eV (Au) to $15.3$ eV (Al). 
\bibitem{Draxl}
P. Puschnig, P. Amiri, and C. Draxl, {Phys. Rev. B} {\bf 86}, 085107 (2012).
\bibitem{Marzari}
Y.S. Lee, M.B. Nardelli, and N. Marzari, {Phys. Rev. Lett.} {\bf 95}, 076804 (2005).
\bibitem{LinChuu}
M.F. Lin and D.S. Chuu, Phys. Rev. B {\bf 56}, 4996 (1997).
\bibitem{FL02}
F. L\'{e}onard, Appl. Phys. Lett. {\bf 81}, 4835 (2002).
\bibitem{Benedict_alpha}
L.X. Benedict, S.G. Louie, and M.L. Cohen, Phys. Rev. B {\bf 52}, 8541 (1995).
\bibitem{Guo}
G.Y. Guo, K.C. Chu, D.-S. Wang, C.-G Duan, Comp. Mat. Sci. {\bf 30}, 269 (2004).
\bibitem{Tamblyn}
I. Tamblyn, P. Darancet, S.Y. Quek, S.A. Bonev, and J.B. Neaton {Phys. Rev. B} \textbf{84}, 201402(R) (2011).
\bibitem{compensation_refs}
Similar compensation effects have also been noticed in the case of semiconductor quantum dots; see C. Delerue et al., Phys. Rev. Lett. {\bf 84}, 2457 (2000) ; A.R. Porter et al., Phys. Rev. B {\bf 64}, 035320 (2001).
\bibitem{Spataru_rad}
C. D. Spataru, S. Ismail-Beigi, R. B. Capaz, and S. G. Louie, Phys. Rev. Lett. {\bf 95}, 247402 (2005).
\bibitem{non-cross}
I assume that the LDA wavefunctions of the CNT-metal system belong either to the CNT or to the metal, neglecting the possibility that states above the vacuum level can be delocalized over both systems. This approximation should introduce little error in the calculated properties of states localized on the CNT. 
\bibitem{Benedict}
Screening in the exchange term was also found to be appropriate in the computation of optical spectra in quantum dots when only a small set of particle-hole states are included in the calculation [see L.X. Benedict, Phys. Rev. B {\bf 66}, 193105 (2002)].
\bibitem{metallicCNT}
In metallic CNTs the free carriers screen out the added charge (the induced charge integrates to the negative of the added charge); this should result in QP renormalization much reduced in comparison to the semiconducting case.
\bibitem{largeR}
For large $R$ one can set $\alpha=a_1 R^2$ in which case $\delta E^{QP}_{model}$ takes the form $\pm\frac{1}{4h_c} f(\frac{R}{h_c}) $,
where the nonanalytic function $f(x)$ is within $\sim10\%$ equal to $1/(1+x)$ for $0<x<1$.
\bibitem{Spataru_ApplPhysA}
C.D. Spataru, S. Ismail-Beigi, L.X. Benedict, S.G. Louie, Appl. Phys. A {\bf 78}, 1129 (2004).
\bibitem{Sohrab}
S. Ismail-Beigi, Phys. Rev. B {\bf 73}, 233103 (2006).

\end{thebibliography}
\end{document}